\documentclass[a4paper,11pt]{amsart}
\begin{document}
\hyphenation{Schwarz-schild ge-ne-ra-ted gra-vi-ta-tio-nal
mea-ning}
\title[The Black Holes do not exist - ``Also Sprach Karl
Schwarzschild''] {{\bf  The Black Holes do not exist\\``Also
Sprach Karl Schwarzschild''}}
\author{Angelo Loinger}
\date{}
\address{Dipartimento di Fisica, Universit\`a di Milano, Via
Celoria, 16 - 20133 Milano (Italy)}
\email{angelo.loinger@mi.infn.it}

\begin{abstract}
According to the \emph{original} theoretical analysis of 1916 by
Karl Schwarzschild the black holes do not have a physical reality.
\end{abstract}

\maketitle

\small \vskip0.25cm\par\hfill {\emph{A paper ``f\"ur Alle und
Keinen''.}}

\normalsize

\vskip1.00cm
\noindent {\bf Introduction}\par \vskip0.10cm
I recall here the
\emph{main} stages of an important theoretical acquisition, which
was originated by a fundamental memoir of 1916 by Karl
Schwarzschild, i.e.: the general theory of relativity (GR) does
\emph{not} allow the physical existence of black holes (BH's) --
if rightly understood. \par From the \emph{observational}
standpoint the alleged discoveries of BH's are mere \emph{flatus
vocis}: in reality, the astronomical observations prove only the
presence of very large -- or enormously large -- masses
concentrated in very small (``punctual'') volumes.

\vskip0.80cm
\noindent {\bf 1916}
\par \vskip0.10cm
In this year Schwarzschild published two epochal memoirs,
respectively entitled ``\"Uber das Gravitationsfeld eines
Massenpunktes nach der EINSTEINschen Theorie'' \cite{1}, and
``\"Uber das Gravitationsfeld einer Kugel aus inkompressibler
Fl\"ussigkeit nach der EINSTEINschen Theorie'' \cite{2}. In the
first paper the Author gives the \emph{exact} solution of the
problem of the Einsteinian gravitational field which is generated
by a point mass $M$ at rest. If $r,\vartheta, \varphi$ are
spherical polar co-ordinates, we have the following expression for
the spacetime interval $\textrm{d}s$:

\begin{equation} \label{eq:one}
    \textrm{d}s^{2}=\left[1-\frac{2m}{\textrm{R}}\right]c^{2}\textrm{d}t^{2}-\left[1-\frac{2m}{\textrm{R}}\right]
    ^{-1}\left(\textrm{d}\textrm{R}\right)^{2}-
    \textrm{R}^{2}\left(\textrm{d}\vartheta^{2}+\sin^{2}\vartheta
    \textrm{d}\varphi^{2}\right),
\end{equation}

where $m\equiv GM/c^{2}$; $G$ is the gravitational constant and
$c$ is the speed of light \emph{in vacuo}; $\textrm{R}\equiv
[r^{3}+(2m)^{3}]^{1/3}$. \par This $\textrm{d}s^{2}$ holds,
physically and mathematically, in the \emph{entire} spacetime,
with the only exception of the origin $r=0$, seat of the mass $M$.
(Remark that this singular point has an associate superficial area
equal to $4\pi(2m)^{2}$: this means simply that the Einsteinian
material point is \emph{not} identical with the Newtonian material
point, as it was emphasized in 1923 by Marcel Brillouin \cite{3}.)
\par In the second paper \cite{2} Schwarzschild determined the
Einsteinian gravitational field generated by an incompressible
fluid sphere. Now, if one computes the limit of this solution when
the sphere contracts into a material point of a finite mass $M$,
one finds anew the Schwarzschildian form of solution for a mass
point of the first memoir (see A. Loinger,
\emph{arXiv:gr-qc/9908009}, August 3rd, 1999). It is very
interesting that a fluid sphere of uniform density and
\emph{given} mass cannot have a radius smaller than $(9/8)(2m)$.

\vskip0.80cm
\noindent {\bf 1916-1917}
\par \vskip0.10cm
If in lieu of the $\textrm{R}$ of eq.(\ref{eq:one}) we put simply
the radial co-ordinate $r$, we obtain the so-called \emph{standard
form} of the solution of the problem solved in \cite{1}
(Schwarzschild problem). This form was discovered independently by
Hilbert \cite{4}, Droste \cite{4}, and Weyl \cite{6}. It is
usually, but \emph{\textbf{erroneously}}, named ``by
Schwarzschild''. The HDW-form is \emph{physically} valid only for
$r>2m$, because within the spatial surface $r=2m$ (a singular
locus) the time co-ordinate takes the role of the radial
co-ordinate, and \emph{vice versa} (and therefore
$\textrm{d}s^{2}$ loses its essential property of physical
appropriateness) and the solution becomes \emph{non}-static. Quite
properly, Nathan Rosen emphasized repeatedly that the radial
co-ordinate of the HDW-form has been initially chosen in such a
way that the area of the space surface $r=k$ is equal to $4\pi
k^{2}$: consequently, it is very difficult to admit that the
co-ordinate $r$ can transform itself into the time co-ordinate
within the space region $r<2m$.
\par Finally, it is very easy to
prove that in the manifold defined by HDW-form for $0<r<\infty$ it
is impossible to assign the \emph{time arrow} to every time
geodesic, according to physically reasonable criteria; on the
contrary, this difficulty does \emph{not} exist for the
Schwarzschild form (\ref{eq:one}), \cite{7}.
\par The fictive notion of BH was generated by \emph{erroneous}
reflections on the ``globe'' $r=2m$ of the standard HDW-form.
\emph{It would not have come forth if the treatises of GR had
expounded the Schwarzschild form of solution in lieu of the
standard form}.
\par Remark that the validity restriction $r>2m$ of the standard form
does \emph{not} imply a \emph{physical} limitation. Indeed, as all
the classic Authors knew, the exterior part $r>2m$ of the HDW-form
is diffeomorphic to the Schwarzschild form, which holds for $r>0$.

\vskip0.80cm
\noindent {\bf 1922-1923-1924}
\par \vskip0.10cm
As far back as 1922 all the competent scientists knew the right
interpretation of the standard HDW-form. Indeed, in 1922 a meeting
was held at the Coll\`ege de France, which was attended by
Einstein. The physical meaning of the ``globe'' $r=2m$ was
discussed and definitively clarified -- see the lucid paper by M.
Brillouin quoted in \cite{3}. This Author investigated also
another interesting form of solution to Schwarzschild problem,
which can be formally obtained by putting in eq.(\ref{eq:one}) the
simple expression $r+2m$ in lieu of R. The validity domain of
Brillouin's form is identical with that of Schwarzschild's form.
Moreover, Brillouin shows that it is \emph{not} permitted to
extend the radial co-ordinate $r$ of Schwarzschild's and
Brillouin's forms to the negative values of the interval
$-2m<r<0$, and proves simultaneously that the attribution of a
physical meaning to the interval $0<r<2m$ of the standard HDW-form
(\emph{as the inventors of the BH's do}) is pure nonsense.
\par In
1924 Eddington published the second edition of his splendid
treatise on Relativity (reprinted in 1930, 1937, 1952, 1954, 1957,
1960), in which we find a very general form of solution to
Schwarzschild problem \cite{8}:

\begin{equation} \label{eq:two}
    \textrm{d}s^{2}=\left[1-\frac{2m}{f(r)}\right]c^{2}\textrm{d}t^{2}-\left[1-\frac{2m}{f(r)}\right]
    ^{-1}\left[\textrm{d}f(r)\right]^{2}-
    \left[ f(r)\right]^{2}\left(\textrm{d}\vartheta^{2}+\sin^{2}\vartheta
    \textrm{d}\varphi^{2}\right),
\end{equation}

where $f(r)$ in \emph{any} regular function of $r$; we see
immediately that: if $f(r)\equiv \textrm{R}$, we have the
Schwarzschild form (\ref{eq:one}); if $f(r)\equiv r$, we have the
standard HDW-form; if $f(r)\equiv r+2m$, we have the Brillouin
form; \emph{etc}. \par \emph{All the physical results are
independent of the particular choice of the function} $f(r)$.
\par
Quite similar considerations can be made for the gravitational
fields generated by \emph{electrically charged} particles.

\vskip0.80cm
\noindent {\bf The years} \emph{\textbf{post}} {\bf 1924}
\par \vskip0.10cm
The previous conception can be easily generalized to the
gravitational field generated by the spinning particle of the
well-known Kerr's solution.
\par In regard to the ``maximally extended'' form of solution to
Schwarzschild problem due to Kruskal \cite{9} and Szekeres
\cite{10} -- a rather baroque form --, we can declare its
\emph{physical superfluity}, because already the (static) forms of
Schwarzschild and Brillouin, in particular, are ``maximally
extended''.
\par \emph{Continued gravitational collapse}: it is almost evident
that if we bear in mind, e.g., Schwarzschild's and Broillouin's
forms, \emph{no continued collapse can generate a BH} -- and this
was just Einstein's opinion \cite{11}.
\par Since 1998 I have published several articles on \emph{arXiv}
in which the non-existence of the BH's has been demonstrated anew.
The papers of the years 1998 $\div$ 2001 have been collected in a
book \cite{12}; papers of the successive years on the same subject
have been published (on \emph{arXiv} and) on \emph{Spacetime and
Substance}.
\par I was motivated by a  simple consideration: \emph{all} the
Great Spirits who founded and developed GR thought that the notion
of BH belongs to science fiction \cite{13}.

\normalsize \vskip0.5cm
\noindent {APPENDIX}
\par \vskip0.10cm
Observations made by a team of astrophysicists of the
\emph{Max-Planck-Institut} for Extraterrestrial Physics have
allowed to determine the positions of the star denote with the
symbol S2 in its motion around the Milky-Way's centre \cite{14}.
It has come out that the S2-orbit is a \emph{Keplerian} ellipse
with a period of 15.2 years.
\par The accuracy of the above research seems indisputable, but
the conclusion of the authors -- according to which the centre
around which S2 revolves is a \emph{black hole} -- seems fully
unjustified. Indeed, the existence of the observed
\emph{Keplerian} orbit can only explain the presence of a
``punctual'' supermassive body at the centre of the Milky-Way --
and not of a supermassive BH \cite{15}.
\par It can be also demonstrated that the two supermassive
celestial bodies at the centre of the distant galaxy NGC 6240 (see
\emph{NASA Press Release}, November 20th, 2002) cannot be black
holes \cite{15}. \\

\end{document}